\documentclass[twocolumn]{jpsj3} 
\usepackage{times}
\def\delx{\partial_x}
\def\d{{\rm d}}

\def\kf{k_{\rm F}}

\def\lsim{\lower -0.3ex \hbox{$<$} \kern -0.75em \lower 0.7ex \hbox{$\sim$}}
\def\im{{\rm i}}
\def\jo #1#2#3#4{#1 {\bf #2} (#3) #4}

\def\PRB{Phys.\ Rev.\ B}
\def\PRL{Phys.\ Rev.\ Lett.}

\def\JPSJ{J.\ Phys.\ Soc.\ Jpn.}

\def\RMP{Rev.\ Mod.\ Phys.}

\def\EPL{Europhys.\ Lett}

\def\CR{Chem.\ Rev.}
\def\EPJB{Eur.\ Phys.\ J.\ B}

\usepackage{bm} 

\def\runauthor{H. Yoshioka \it{et al.}}
\title{
Finite-Temperature Properties across the Charge Ordering Transition\\
-- Combined Bosonization, Renormalization Group, and Numerical Methods --
}

\author{%
Hideo \textsc{Yoshioka}%
  $^1$\thanks{E-mail: h-yoshi@cc.nara-wu.ac.jp},
Masahisa \textsc{Tsuchiizu}%
  $^2$\thanks{E-mail: tsuchiiz@s.phys.nagoya-u.ac.jp}, 
Yuichi \textsc{Otsuka}%
  $^{3,4}$\thanks{E-mail: otsukay@riken.jp}, 
and
Hitoshi \textsc{Seo}%
  $^{3,4}$\thanks{E-mail: seo@riken.jp}
}

\inst{%
$^1$Department of Physics, Nara Women's University, Nara 630-8506 \\
$^2$Department of Physics, Nagoya University, Nagoya 464-8602 \\
$^3$Condensed Matter Theory Laboratory, RIKEN, Wako, Saitama 351-0198 \\
$^4$JST, CREST, Saitama 351-0198
}

\recdate{\today}

\abst{
We theoretically describe the charge ordering (CO) metal-insulator transition 
 based on a quasi-one-dimensional extended Hubbard model, 
 and investigate the finite temperature ($T$) properties across 
 the transition temperature, $T_{\rm CO}$. 
 In order to calculate $T$ dependence of physical quantities 
  such as the spin susceptibility and the electrical resistivity,  
  both above and below $T_{\rm CO}$, 
a theoretical scheme is developed 
 which combines analytical methods 
 with numerical calculations. 
We take advantage of the renormalization group equations 
 derived from the effective bosonized Hamiltonian, 
 where Lanczos exact diagonalization data are chosen as initial parameters, 
 while the CO order parameter at finite-$T$ 
 is determined by quantum Monte Carlo simulations.  
The results show that the spin susceptibility does not show 
 a steep singularity at $T_{\rm CO}$, and it slightly increases 
 compared to the case without CO because of the suppression of 
the spin velocity. 
In contrast, the resistivity 
 exhibits a sudden increase at $T_{\rm CO}$, below which 
 a characteristic $T$ dependence is observed. 
We also compare our results with experiments on 
 molecular conductors as well as transition metal oxides
 showing CO. 
}

\kword{charge ordering, molecular conductors, 
 extended Hubbard model,
 exact diagonalization, bosonization, renormalization group, 
 quantum Monte Carlo, transition metal oxides}

\begin{document}
\maketitle

\section{Introduction} 
\label{sec.Introduction}

Charge ordering (CO) phase transition is now found ubiquitously 
 in strongly correlated electron systems such as 
 transition metal oxides~\cite{Imada1998} and 
 molecular conductors~\cite{Seo2004}. 
Its intuitive picture is simple: 
 Electrons arrange spontaneously which results in lowering of the symmetry from  
 the underlying lattice structure, 
 in order to gain repulsive Coulomb energy as in Wigner crystals. 
Nevertheless, the richness of this phenomenon is now widely recognized, 
 owing to extensive experimental as well as theoretical investigations 
 in many types of compounds with different lattice geometries. 

From the theoretical point of view, in spite of numerous studies~\cite{Seo2006}, 
there remains a fundamental question not fully clarified yet, 
 i.e., how the physical properties across the 
 CO phase transition temperature ($T$), $T_{\rm CO}$, can be described. 
Typical mean-field analysis 
 fails in reproducing finite-$T$ phase transitions 
 from a metallic state to 
 paramagnetic insulating CO states, 
 which are, however, 
 often observed in many strongly correlated electronic materials. 
Therefore, treatments beyond the simple mean-field approximation, 
 which consider the effects of quantum fluctuation more properly, are necessary 
 to describe such properties. 
  
\begin{figure}
\begin{center}
\includegraphics[width=7cm]{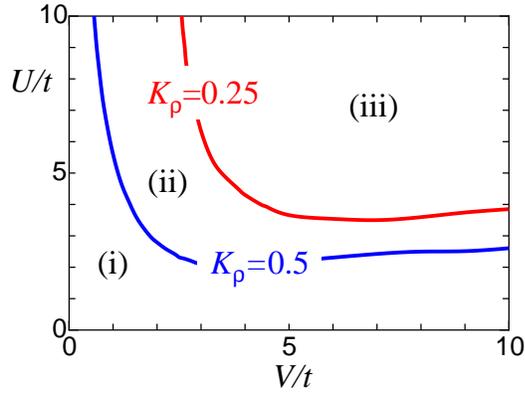}
\end{center}
\caption{
(Color online)
Phase diagram of the quasi-one-dimensional extended Hubbard model,
 obtained by the interchain mean-field theory~\cite{YTS2006}. 
In region (i), the CO state is stabilized  
  when $V_\perp>V_\perp^{\mathrm{c}}$ with finite 
  $V_\perp^{\mathrm{c}}>0$. 
In region (ii),  infinitesimal $V_\perp$ changes the TL liquid
 state at $V_\perp = 0$ into the CO state. 
In region (iii), where the CO state is obtained even in the purely 1D
 case, the CO state is obtained at finite temperature 
by the  interchain coupling.
The boundary between (i) and (ii), and that between (ii) and (iii)
are characterized by the value of $K_\rho = 1/2$ and that of 
$K_\rho = 1/4$ for the 1D model.}
\label{f2}
\end{figure}

In general, one-dimensional (1D) models can be treated by 
 taking the quantum fluctuations into account in a controlled way, 
 compared to higher dimensional systems, 
 by numerical as well as analytical methods. 
In fact, to describe CO in quarter-filled systems, 
the 1D extended Hubbard model (EHM) 
 including the repulsive Coulomb interactions of onsite, $U$, and intersite, $V$, 
 has intensively been studied. 
Especially, its ground-state properties (see Fig.~\ref{f2})
 are known in detail;  
 a quantum phase transition occurs between the metallic 
 Tomonaga-Luttinger (TL) liquid state characterized by the TL liquid parameter 
  $K_\rho$, and the CO insulating state ($T_{\rm CO}=0$)~\cite{Seo2006,YTS2000,Ejima2005}. 
However, the 1D model 
 does not show any phase transition at finite $T$ 
 due to the enhanced low-dimensional fluctuations.

On the other hand, 
 the two-dimensional (2D) square lattice EHM at quarter-filling 
 shows finite $T_{\rm CO}$. 
Different techniques beyond the mean-field approximation have been applied 
 to investigate the finite-$T$ properties of this model, 
 such as 
 exact diagonalization (ED)~\cite{Hellberg2001}, 
 slave-boson~\cite{Merino2001}, 
 dynamical mean-field~\cite{Pietig1999,Merino2006}, 
 correlator projection~\cite{Hanasaki2005}, 
 and quantum Monte Carlo (QMC)~\cite{Tanaka2006} methods. 
However, 
 due to theoretical difficulties, 
 the physical properties across $T_{\rm CO}$ 
 such as the spin susceptibility 
 and the electrical resistivity are not elucidated, 
 and the interplay 
 between spin and charge degrees of freedom is not fully explored yet. 

In this context, 
 the quasi-1D (Q1D) EHM, i.e., 
 1D EHM chains coupled by the interchain Coulomb interaction $V_\perp$, 
 has recently been studied by analytical~\cite{YTS2006,YTS2007}
 as well as numerical~\cite{Seo2007,Otsuka2008} methods  
 by the present authors and co-workers, 
 which shows a finite-$T$ CO phase transition with concomitant metal-insulator transition 
 at quarter-filling.
In these studies, the interchain mean-field treatment~\cite{Scalapino1975} 
 is applied and the resultant effective 1D model is solved using different methods 
 which properly take into account the large fluctuation effects: 
 by the bosonization + renormalization group (RG) 
 scheme in refs.~\citen{YTS2006} and \citen{YTS2007}, 
 and by numerical techniques, i.e., 
the quantum transfer-matrix method 
 in ref.~\citen{Seo2007} and the QMC method
 in ref. \citen{Otsuka2008}. 

In the former analytical approach, 
 which has the advantage in investigating the critical regions, 
 it is found that the $V_\perp$-term  
 considerably affects the stability of the CO state 
 compared with that in the 1D EHM~\cite{YTS2006}. 
Due to the dimensionality effect, 
 $T_{\mathrm{CO}}$ always becomes finite 
 whenever the CO phase is realized, 
 except for the critical point (line). 
Then, the ground state phase diagram 
 of the 1D EHM on the $U$-$V$ plane 
 is divided 
 into three regions depending on the TL liquid parameter $K_\rho$, 
 as in Fig. \ref{f2}. 
They show different properties when $V_\perp$ is turned on: 
\begin{itemize}
\item 
Region (i) [$K_\rho > 1/2$]
 Finite value of $V_\perp$ is necessary to produce the CO state. 
\item Region (ii) [$1/4 \leq K_\rho \leq 1/2$] 
 Infinitesimal $V_\perp$ turns the system from TL liquid to CO insulator
      with finite $T_{\mathrm{CO}}$. 
\item Region (iii) [$K_\rho$ is not defined (CO insulating ground state for $V_\perp =0$)] 
 Infinitesimal $V_\perp$ makes $T_{\mathrm{CO}}$ finite. 
\end{itemize} 
Therefore, once $V_\perp$ is added, 
 the CO phase critically enlarges from region (iii) to regions (ii)+(iii). 
In addition, the effects of the lattice dimerization along the chain direction
 and the frustration in the interchain interactions
 on the above Q1D model have been studied~\cite{YTS2007}; 
the lattice dimerization suppresses $T_{\mathrm{CO}}$, 
 and 
 the interchain frustration 
 leads to a competition between CO states
 with different charge patterns~\cite{Seo2001}. 

In these studies, 
 although the finite-$T$ properties above $T_{\rm CO}$ 
 are elucidated by taking advantage of the RG treatment 
 which we will describe later, 
 properties {\it below} $T_{\rm CO}$ were hardly investigated  
 due to difficulties in determining the CO order parameter 
 in a self-consistent manner. 
Meanwhile, 
 the `initial' values in the RG equations were 
 taken as the bare TL parameters 
 derived from weak-coupling expansions of the EHM; 
 such a procedure loses accuracy in general when the interaction becomes large. 
Nevertheless, 
 this drawback is not due to 
 the phase Hamiltonian representation 
 by the bosonization method nor the RG approach;
 it is because of the choice of the initial condition for the RG equations.   
 
On the other hand, 
 the numerical quantum transfer matrix and QMC methods 
 used in refs. \citen{Seo2007} and \citen{Otsuka2008} 
 are suitable for the intermediate-to-strong coupling regime; 
 by applying the interchain mean-field approach as well, 
 the Q1D EHM and its extensions with different types of electron-lattice interaction 
 are treated. 
Competitions and co-existences among different 
 states, including the paramagnetic CO insulator, are clarified, 
 and the $T$ dependences of different order parameters 
 as well as the charge and spin susceptibilities 
 across  the transition temperatures are computed. 
In particular, the QMC method could provide highly accurate results down 
 to considerably low $T$. 
However, there are quantities 
 which cannot be calculated by such numerical simulations directly, 
 such as the electrical resistivity: 
 one of the most basic information from experiments.

Up to now, the above-mentioned analytical and numerical approaches 
 for the Q1D quarter-filled electron systems 
 have been separately employed, 
 being complimentary to each other. 
In the present study, in contrast,
 we develop a combined method. 
The numerical ED data  
 are implemented 
 into the analytical bosonization + RG scheme, 
 as initial conditions of the RG equations~\cite{Sano2004}. 
As for the properties below $T_\mathrm{CO}$, 
 the $T$ dependence of the CO order parameter 
 is calculated by the QMC method~\cite{Otsuka2008}, 
  and then adopted to the scheme above. 
This combined method enables us 
 to compute the $T$ dependences of 
 physical properties such as 
 the spin susceptibility and the electrical resistivity 
{\it across} $T_\mathrm{CO}$. 

The organization of this paper is as follows. 
In \S\ref{sec.Formulation}, 
 our combined analytical and numerical method 
 applied to the Q1D EHM is formulated. 
In \S\ref{sec.T-dep}, 
 the $T$ dependences of the physical quantities 
 across the CO transition 
 are shown.  
Section \ref{sec.Discussions} is devoted to the summary and discussions, 
 including comparisons with experiments on different CO materials. 
Detailed description of our theoretical approach is presented in 
 Appendix \ref{ap.combined_approach}, 
 and a benchmark check of this method applied to the 1D Hubbard model 
 is given in Appendix \ref{ap.susceptibility}.

\section{Formulation}
\label{sec.Formulation}

In this section, the model and formulation for our calculation are given. 
In \S~\ref{subsec.Bosonization+RG}, 
 first we explain the interchain mean-field approach to the Q1D EHM 
 and the bosonization + RG method applied to the effective 1D model, 
 which were partly formulated 
 in refs.~\citen{YTS2006} and \citen{YTS2007}. 
 Then, in \S~\ref{subsec.Numerical} 
 we explain how the numerical techniques are combined with this method. 

\subsection{Bosonization + RG}
\label{subsec.Bosonization+RG}

The Q1D EHM that we study consists of
 quarter-filled 1D extended Hubbard chains
 coupled by the interchain interaction $V_\perp$.~\cite{YTS2006}  
The Hamiltonian $H_{\mathrm{Q1D}}$ is given by
 \begin{align}
H_{\mathrm{Q1D}} &= \sum_j H^{j}_{\mathrm{1D}} + H_\perp, \label{eqn:HT} \\ 
H^{j}_{\mathrm{1D}} =&
-t \sum_{i,s}  
   \left( c^\dagger_{i,j,s} c_{i+1,j,s}^{} + \mathrm{h.c.} \right) 
\nonumber \\ &
+ U \sum_{i}  n_{i,j,\uparrow} \, n_{i,j,\downarrow}
+ V \sum_{i} n_{i,j} \, n_{i+1,j} , 
\label{eqn:H1dEHM}\\
H_\perp 
&
= V_\perp \sum_{i,\langle j,j' \rangle} n_{i,j} \, n_{i,j'}, 
\label{eqn:Hperp}
\end{align}
where $H^j_{\mathrm{1D}}$ and $H_\perp$ represent the 1D EHM 
of the $j$-th chain and the interchain-coupling term, 
respectively. 
Here, $t$ is the transfer integral
 between the nearest-neighbor sites along the chain direction; 
 we do not consider the interchain hopping energy here. 
The operator $c^\dagger_{i,j,s}$ creates an electron
 with spin $s=\uparrow$~or~$\downarrow$ at the $i$-th site
 in the $j$-th chain.
The density operators are defined as
$n_{i,j,s} = c^\dagger_{i,j,s} c_{i,j,s}^{}-1/4$ and 
$n_{i,j} = n_{i,j,\uparrow} + n_{i,j,\downarrow}$.

The interchain coupling term is treated within the
interchain mean-field approximation~\cite{Scalapino1975} as 
\begin{equation}
 H_\bot 
\simeq  V_\bot \sum_{i,\langle j,j'\rangle} 
\left(
\langle n_{i,j} \rangle n_{i,j'} + n_{i,j} \langle n_{i,j'}
 \rangle 
- 
\langle n_{i,j} \rangle \langle n_{i,j'} \rangle 
\right).
\end{equation}  
Assuming the Wigner-crystal-type CO with 
  two-fold periodicity, 
 we postulate $\langle n_{i,j} \rangle = (-1)^{i+j} n$
   where $n$ is the CO order parameter~\cite{YTS2006}.
After this procedure, 
  we obtain an effective 1D Hamiltonian:
 \begin{align}
H =&
-t \sum_{i,s}  
   \left( c^\dagger_{i,s} c_{i+1,s}^{} + \mathrm{h.c.} \right) 
\nonumber \\ &
 + U \sum_{i}  n_{i,\uparrow} \,n_{i,\downarrow}
 + V \sum_{i} n_{i} \, n_{i+1}  
\nonumber \\
&+ z V_\perp n \sum_{i} (-1)^i n_{i}
 + \frac{1}{2} z L V_\bot n^2,
\label{eqn:5} 
\end{align}
where the chain index is omitted. 
The number of the sites in the chain 
 and
that of the adjacent chains 
are expressed as $L$ and $z$, respectively. 

By the bosonization method,
  the effective 1D Hamiltonian eq. (\ref{eqn:5}) 
  can be expressed in terms of the bosonic fields. 
Then the Hamiltonian is separated into 
  the charge part $\mathcal{H}_\rho$ 
  and the spin part $\mathcal{H}_\sigma$ as 
\begin{align}
H= & \int \d x(\mathcal{H}_\rho+\mathcal{H}_\sigma) + \frac{1}{2} z L V_\bot n^2, \\
 \mathcal{H}_\rho 
=
& \frac{v_{\rho0}}{4\pi} 
\left[
\frac{1}{K_{\rho0}} (\delx \theta_\rho)^2 + K_{\rho0} (\delx \phi_\rho)^2
\right]
\nonumber \\
&
+\frac{g_{3\bot}}{(\pi \alpha)^2} \cos 2 \theta_\rho 
 +\frac{g_{1/4}}{2(\pi \alpha)^2}  \cos 4 \theta_\rho,  
\label{eqn:Hrho}
\\
 \mathcal{H}_\sigma =& \frac{v_{\sigma0}}{4\pi} 
\left[
\frac{1}{K_{\sigma0}} (\delx \theta_\sigma)^2 + K_{\sigma0} (\delx \phi_\sigma)^2 
\right]
\nonumber \\
&+\frac{g_{1\bot}}{(\pi \alpha)^2} 
 \cos 2\theta_\sigma,   
\label{eqn:Hsigma}
\end{align} 
where the phase variables satisfy
 $[\theta_\mu(x),\phi_\nu(x')] = \im \pi \mathrm{sgn} (x-x') 
 \delta_{\mu \nu}$ with $\mu, \nu = \rho$ or $\sigma$, 
 and $\alpha$ is a short-distance cutoff. 
The parameters $K_{\rho0}$ ($K_{\sigma0}$) and $v_{\rho0}$
  ($v_{\sigma0}$)  are the 
  \textit{bare} TL-liquid
  parameter and velocity for the charge (spin) degree of 
 freedom, respectively.
They take non-universal values depending on the strength of 
  interactions.

In the charge sector, $\mathcal{H}_\rho$, 
there appear two kinds of non-linear terms. 
The term proportional to $\cos 4 \theta_\rho$ 
 originates from the $8\kf$ umklapp scattering
 [$\kf(=\pi/4)$ is the Fermi momentum], 
whose coupling constant $g_{1/4}$ is finite even in the purely 1D EHM 
 and it leads to the CO insulating ground state~\cite{YTS2000,Schulz1994,Giamarchi1997}. 
On the other hand, 
 the $\cos 2 \theta_\rho$ term represents 
 the $4\kf$ umklapp scattering process,
 whose coupling constant $g_{3\perp}$ 
 is proportional to $n$ when $|g_{3\bot}| \ll \pi v_{\rho 0}$~\cite{YTS2006}. 
This $4\kf$ umklapp scattering process
 is generated in the existence of CO, 
 which can be understood by noting that
 the one-particle energy gap
 opens at $\pm 2 \kf$ 
 because of the two-fold periodicity in the CO state
 and then
 the conduction band becomes effectively half filled.

The spin part $\mathcal{H}_\sigma$ is essentially the same as 
  the effective Hamiltonian of the Heisenberg chain.
The parameters $K_{\sigma0}$ and $g_{1\bot}$ 
  in eq.\ (\ref{eqn:Hsigma}) are not independent 
  because of the spin-rotational SU(2) symmetry: 
\begin{align}
 K_{\sigma0} 
= \sqrt{\frac{\pi v_{\sigma0} +g_{1\bot}}{\pi v_{\sigma0}-g_{1\bot}}}. 
\label{eqn:SU2}
\end{align} 
This constraint still holds even under the scaling procedure. 
In the low-energy limit, the $g_{1\perp}$ coupling is renormalized to 
  zero and $K_{\sigma0}$ reduces to unity.
When the system has the SU(2) symmetry,
 it is known that  physical quantities exhibit logarithmic
 (very slow) system-size or $T$ dependences due to 
 the presence of marginal operators.\cite{Cardy}
These characteristics can be captured by the RG analysis.

The RG equations 
for 
 the charge part $\mathcal{H}_\rho$ [eq.\ (\ref{eqn:Hrho})] 
 are given by 
\begin{align}
 \frac{\d}{\d l} K_\rho (l)
&=
 -2 G_{3\bot}^2(l) \, K_\rho^2(l)
 - 8 G_{1/4}^2(l) \, K_\rho^2(l), 
\label{eqn:RGC1}
\\
 \frac{\d}{\d l} G_{3 \bot}(l) 
&=
 (2 - 2K_\rho(l)) G_{3 \bot}(l) -
 G_{3\bot}(l) \, G_{1/4}(l), 
\label{eqn:RGC2}
\\
 \frac{\d}{\d l} G_{1/4} (l)
&= 
(2 - 8K_\rho(l))G_{1/4}(l) -
 \frac{1}{2}G_{3\bot}^2(l).  
\label{eqn:RGC3}
\end{align}
We note that, 
 in ref.~\citen{YTS2006}, 
 the above equations with $G_{3\bot}(l)=0$  ($g_{3 \bot} = 0$)  were treated, 
 corresponding to the absence of CO order parameter $n$,  
 to investigate the instability toward $T_{\rm CO}$. 
As for the spin part, the scaling equation for the coupling in 
 $\mathcal{H}_\sigma$ [eq. (\ref{eqn:Hsigma})] is
\begin{align}
 \frac{\d}{\d l} G_{1\bot}(l)
 = - 2 G_{1\bot}^2(l) - 2 G_{1\bot}^3(l),
\label{eqn:RGS}  
\end{align}
and $K_{\sigma}(l)$ is determined as 
 \begin{align}
K_\sigma(l)= \sqrt{\frac{1+G_{1\bot}(l)}{1-G_{1\bot}(l)}},
\label{eqn:SU22}
\end{align}
following eq. (\ref{eqn:SU2}). 
In the usual analysis, 
 to obtain the parameters $\{K_\rho (l), G_{3 \bot}(l) , G_{1/4} (l)\}$ for the charge part 
 and $G_{1\bot}(l)$ for the spin part as the solutions of the RG equations,  
 the initial conditions are set from the bare parameter values as 
 $K_{\rho}(0)=K_{\rho0}$,
 $G_{3\perp}(0)=g_{3\perp}/(\pi v_{\rho0})$,
 $G_{1/4}(0)=g_{1/4}/(2\pi v_{\rho0})$, and
 $G_{1\perp}(0)=g_{1\perp}/(\pi v_{\sigma0})$. 
 Such initial values can be calculated from 
 the parameters of the original lattice model 
 by considering the interaction processes near the Fermi level. 
In our previous studies on CO based on such 
 procedure~\cite{YTS2000,Tsuchiizu2001,YTS2006,YTS2007}, 
 the third-order processes mediated by the states far from the Fermi level
 were crucial in deriving the $8 \kf$ umklapp scattering $g_{1/4}$-term, 
 which triggers the CO insulating state.  
Note that the third-order virtual processes also play crucial roles 
for the spin degree of freedom~\cite{YTS2000}.

We also note that the RG equations above 
 for the charge part are
 shown up to the one-loop level, while, on the other hand, 
 that for the spin coupling $G_{1\perp}(l)$ is shown 
 up to the two-loop level, i.e., $O(G_{1\perp}^3)$.
Since there are subtleties in deriving the two-loop RG equation based on 
 the bosonized Hamiltonian, 
 we follow the consideration given in ref. \citen{Emery} and 
  use the RG equation for $G_{1\perp}$
  based on the Hamiltonian for the original fermion variables.

When we apply the RG method to systems at finite $T$, 
 the assumption of the scaling invariance breaks down 
  and the RG scaling is cut off at 
  the scale $l$ corresponding to the temperature $T$: 
 $l \simeq l_T \equiv \ln(Ct/T)$ with $C$
 being an $O(1)$ numerical constant. 
Then we can discuss the finite-$T$
 properties by terminating the scaling procedure at $l_T$, 
 and use the values 
 $\{K_\rho (l_T), G_{3 \bot}(l_T) , G_{1/4} (l_T)\}$ and $G_{1\bot}(l_T)$ 
 as the $T$-dependent quantities; 
 we write them as $\{K_\rho (T), G_{3 \bot}(T) , G_{1/4} (T)\}$ and $G_{1\bot}(T)$ 
 in the following. 
These $T$-dependent parameters are set in  
 the formulae for the physical quantities,  
 as will be discussed in \S~\ref{sec.T-dep}.

\subsection{Numerical methods}
\label{subsec.Numerical}

The procedure of setting the initial conditions 
 of the RG equation mentioned above, 
 i.e., to derive them based on the perturbation theory, 
 generally looses accuracy in the strong-coupling region. 
For example, 
 the ground state phase diagram of the 1D EHM 
 obtained in this way 
 qualitatively agrees with the other numerical methods 
 in the weak-coupling region,
 while the phase boundary between TL liquid and CO insulator 
 deviates at strong-coupling~\cite{YTS2000}. 
In the present study, instead, 
 in order to obtain more accurate results even for stronger interactions, 
 we make use of numerical data for finite size systems, as discussed in
 the following. 
Such an approach was recently proposed in ref.~\citen{Sano2004} 
 to investigate the charge degree of freedom 
 of the 1D EHM, 
 which reproduced the ground-state phase diagram with high accuracy. 
Here we extend their method 
 in order to calculate the finite-$T$ properties 
 by solving the RG equations in \S~\ref{subsec.Bosonization+RG} 
 and stopping the scaling at $l_T$ as mentioned above. 

For small systems with sites $L$, 
 the TL parameters $K_\rho^{L}$ and $K_\sigma^{L}$ 
 can be computed by the Lanczos ED technique 
 directly applied to the lattice Hamiltonian eq.~(\ref{eqn:5}). 
These can be used as initial values of the RG equations 
 by taking advantage of the relation between 
 the scaling variable $l$ and the system size, 
 $l \simeq \ln{L}$ (in the calculation we use $l = \ln{L}$ since 
the deviation is small for a different factor ).  
For the charge part, 
 eqs. (\ref{eqn:RGC1})-(\ref{eqn:RGC3}), 
 where the three parameters  
  $\{K_\rho (l), G_{3 \bot}(l) , G_{1/4} (l)\}$
  are coupled in the form of differential equations, 
  we can set $K_\rho^{L}$ for available system sizes as initial values 
 by fitting them to the RG equations to determine the solutions. 
As for the spin degree of freedom, 
  the initial value for $G_{1\bot}(l)$ 
  in eq. (\ref{eqn:RGS}) is determined from 
  $K_\sigma^{L}$ using the relation eq.~(\ref{eqn:SU2}). 
It will be discussed in \S~\ref{subsec.uniform_susceptibility} that, 
 for the calculation of the spin susceptibility, 
 we also make use of $T$-dependent spin velocity $v_\sigma(T)$. 
However, the spin velocity cannot be determined reliably from the RG flow 
 due to subtleties in deriving its RG equation.
Therefore, instead, we use the standard polynomial finite size scaling 
 for several system sizes $L$
 and extrapolate it to the size corresponding to $\ln{L} \simeq l_T$, 
 which provides $v_\sigma(T)$. 
More detailed procedure for determining 
 the $T$-dependent parameters 
 using the Lanczos ED data is given in Appendix A. 

We also combine numerical data for the CO order parameter $n$. 
It is finite for $T<T_{\rm CO}$, 
 which affects the parameters of the system; 
 especially $g_{3 \bot}$ ($G_{3 \bot}$) becomes finite when $n$ $\neq$ 0. 
However, its $T$ dependence cannot be obtained 
 within the bosonization + RG scheme 
 in a self-consistent manner, 
 due to the ambiguity 
in the relationship 
 between $n$ and 
 the phase variable~\cite{note}. 
In order to overcome this difficulty,  
 in the present study,  
 $n$ is numerically obtained by the QMC method in advance,  
 and adopted to the above scheme. 
We employ the stochastic-series-expansion (SSE) method~\cite{AWS_1991,AWS_1992}
 with the operator-loop update~\cite{AWS_1999,Sengupta_2002}
 as a solver for the effective 1D lattice Hamiltonian (\ref{eqn:5}).
At each $T$, 
 the order parameter is iteratively calculated 
 by this QMC method 
 until it converges~\cite{Otsuka2008}. 
We calculate systems with sizes up to $L=128$ sites 
 and checked that
 finite size effects are negligible down to 
low $T$ that we discuss in this paper.  
The results are substituted 
 into eq.~(\ref{eqn:5}) and   
 treated as an `external field', 
 which reflect the initial values of the RG equations. 

Summarizing, 
 at each $T$, 
 first the value of $n$ is calculated by QMC, 
 and then the initial conditions of the RG equations 
 are collected using Lanczos ED 
 by substituting the QMC data into the lattice Hamiltonian, 
 and finally the RG equations (and the usual finite size scaling for the spin velocity) 
 are solved by 
 stopping the scaling at the scale $l$ corresponding to the temperature $T$. 
This provides the finite-$T$ values of the parameters 
 to be input in the expressions of the physical quantities 
 which will be described in the next section.  

\section{Temperature Dependences of Spin Susceptibility and Electrical Resistivity}
\label{sec.T-dep}

In this section, the $T$ dependences  
 of the spin susceptibility $\chi_\sigma(T)$ and the electrical resistivity $\rho (T)$
 are discussed. 
We focus on the regions (ii) and (iii) in  Fig.\ \ref{f2} and 
 study how the physical quantities behave for wide $T$ ranges 
  across $T_{\rm CO}$. 
For the intrachain parameter, 
 we set 
 $(U/t,V/t)=(6.0,\, 2.5)$ for the region (ii) 
 and
 $(10.0,\, 4.0)$ for the region (iii). 

\subsection{Uniform spin susceptibility}
\label{subsec.uniform_susceptibility}

The formula for the 
 uniform spin susceptibility $\chi_\sigma(T)$ in the RG scheme  
 has been derived 
 in refs. \citen{Nelisse1999} and \citen{Fuseya2005}. 
The naive random-phase-approximation (RPA) 
 gives 
 $\chi_\sigma^{\mathrm{RPA}} (T) = [2\chi_0(T) /{\pi v_F}] 
/ [ 1 - U \chi_0(T)/(\pi v_F)]$,  
 where 
 $v_F$ is the Fermi velocity and
 $\chi_0(T)$ is the spin susceptibility in the non-interacting case 
 normalized as $\chi_0(0) = 1$.  
On the other hand, 
 when the 1D fluctuation effects are taken into account 
 by the RG method, 
 it is written as~\cite{Nelisse1999,Fuseya2005}, 
\begin{align}
 \chi_\sigma (T) = \frac{2}{\pi v_F} 
\frac{\chi_0(T)}
{1 - \left[G_{1\bot}(T) + G_{4\sigma}(T) \right] \chi_0(T)}. 
\label{eqn:chis}
\end{align} 
Here, $G_{1\bot}(T)$ is
 the amplitude of the backward scattering, 
 which is obtained by the RG scheme in \S~\ref{sec.Formulation}. 
The coupling $G_{4\sigma}(T)$,  on the other hand, 
  represents 
  the same-branch forward scattering in the spin channel. 
  \cite{Nelisse1999}. 
It reflects the velocity of the spin excitations through the relation 
\begin{align}
 G_{4\sigma}(T) = 1 - \frac{v_\sigma(T)}{v_F},  
\label{eqn:g4s}
\end{align}
 then we can use the $T$-dependent spin velocity $v_\sigma(T)$ 
 obtained by the finite size scaling introduced in \S~\ref{sec.Formulation}. 
We note that eqs. (\ref{eqn:chis}) and (\ref{eqn:g4s})  
  reproduce the correct formula 
  $\chi_\sigma(0) = 2/[\pi v_\sigma(0)]$
  in the $T\to 0$ limit.
In refs. \citen{Nelisse1999} and \citen{Fuseya2005}, 
the linearized dispersion was used in deriving 
 the noninteracting susceptibility $\chi_0(T)$.
Here, instead, in order to analyze $\chi_\sigma (T)$
 in a wider $T$ range, 
 we use the dispersion of the tight-binding model 
 to calculate $\chi_0(T)$ (see also Appendix B). 

\begin{figure}[t]
\begin{center}
\includegraphics[width=6cm]{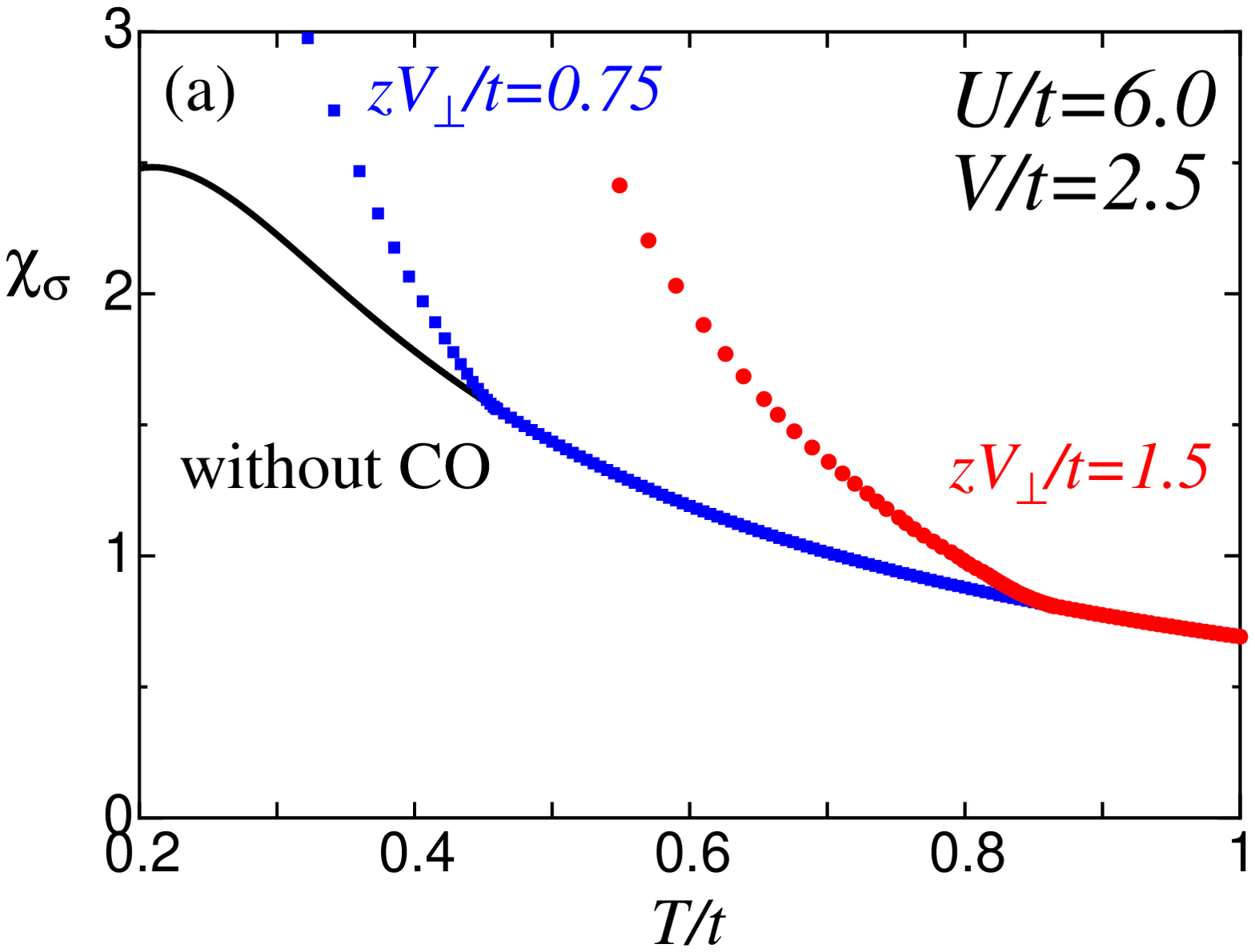} \\
\includegraphics[width=6cm]{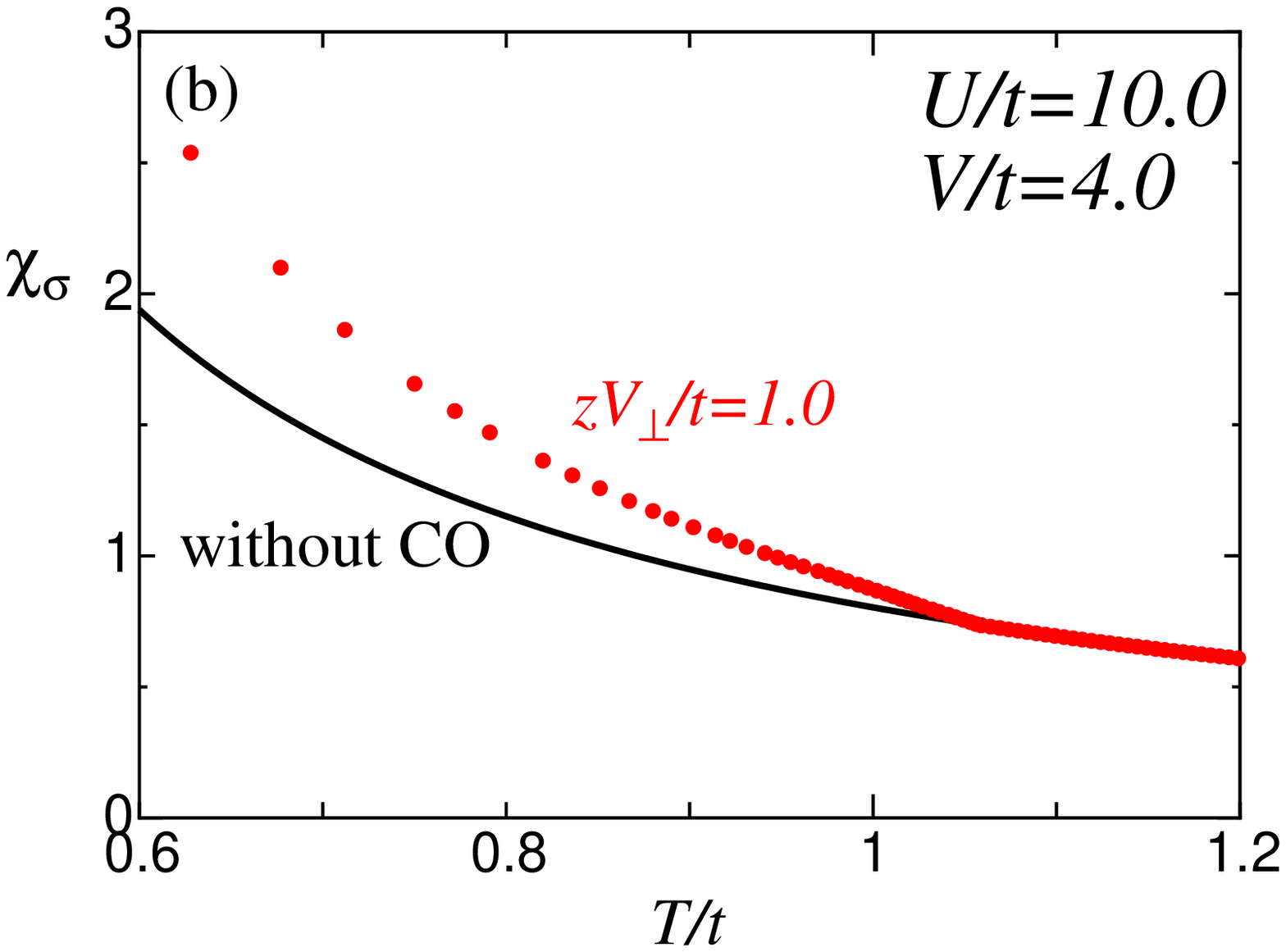}
\end{center}
\caption{
(Color online)
$T$ dependence of the spin susceptibility $\chi_\sigma(T)$ 
 (in the unit of  $1/(ta)$), 
(a) for the region (ii) in Fig. \ref{f2}[$(U/t,V/t)=(6.0,\, 2.5)$]
  and 
(b) for the region (iii) [$(U/t,V/t)=(10.0,\, 4.0)$]. 
The solid curve shows the data for $V_{\perp}=0$, 
 without charge order at finite $T$.
}
\label{f3}
\end{figure}

In Fig. \ref{f3},
 the results are shown 
 for regions (ii)  and  (iii),  
 where, in both cases, 
$\chi_\sigma(T)$ is enhanced below $T_{\rm CO}$, 
 without any steep singularity at $T = T_{\rm CO}$. 
The main reason for the enhancement is as follows. 
The CO order parameter $n$ 
 induces a gap formation at $k= \pm 2 \kf = \pm \pi/2$ 
 in the energy dispersion, 
 and hence 
 the density of states at 
 the Fermi energy is increased
 (the Fermi velocity is suppressed). 
This leads to the suppression of the spin velocity. 

The same approach can be applied to 
evaluate the splitting of Knight shift, 
which is often observed in experiments
to detect the CO transition. 
In the interchain mean-field approach, 
the Knight shift at the charge rich site and the poor site,
$S_+$ and $S_-$, are given by \cite{YTS2007} 
\begin{eqnarray}
 S_{\pm} \propto \chi_\sigma(T) 
\left[ 1 \pm \frac{z V_\perp n}{\sqrt{2t^2 + (z V_\perp n)^2}}
\right]. 
\label{eq:shift}
\end{eqnarray}
For small $V_\perp n/t$, we obtain a simple relation 
 $ S_{\pm} \propto \chi_\sigma(T) 
  [ 1 \pm z V_\perp n/ 2t ] $.
Similarly,   
 we can derive the 
 formula of the nuclear relaxation rate, 
whose separation is also an experimental evidence of CO:    
\begin{eqnarray}
 \frac{1}{(T_{1 \pm}) T} 
\propto 
\left[ 1 \pm \frac{z V_\perp n}{\sqrt{2t^2 + (z V_\perp n)^2}}
\right]^2,  
\label{eq:rate}
\end{eqnarray}  
where the relaxation rate at the charge rich and poor sites 
 are denoted by $(T_{1+})^{-1}$ and $(T_{1-})^{-1}$,
respectively. 
Note that the discrepancy in the order of the factor 
$1 \pm z V_\perp n/ \sqrt{2t^2 +(z V_\perp n)^2}$ 
between (\ref{eq:shift}) and (\ref{eq:rate})  
comes from the fact that 
the former expresses the local response under the {\it uniform}
 perturbation, whereas the latter is the local response under the {\it local}
perturbation. 
Namely, $S_i \propto \sum_{i'} \chi_\sigma (i,i';0)$ and 
$(T_{1i} T)^{-1} \propto \lim_{\omega \to 0} {\rm Im} \chi_\sigma
 (i,i;\omega)/\omega$, 
where $S_i$ and $T_{1i}^{-1}$ are the Knight shift and the NMR
 relaxation rate at the $i$-th site, 
and $\chi_\sigma (i,i';\omega)$ is the dynamical spin
 susceptibility in the site representation.

\subsection{Electrical resistivity}
\label{subsec.resistivity}

\begin{figure}[t]
\begin{center}
\includegraphics[width=6cm]{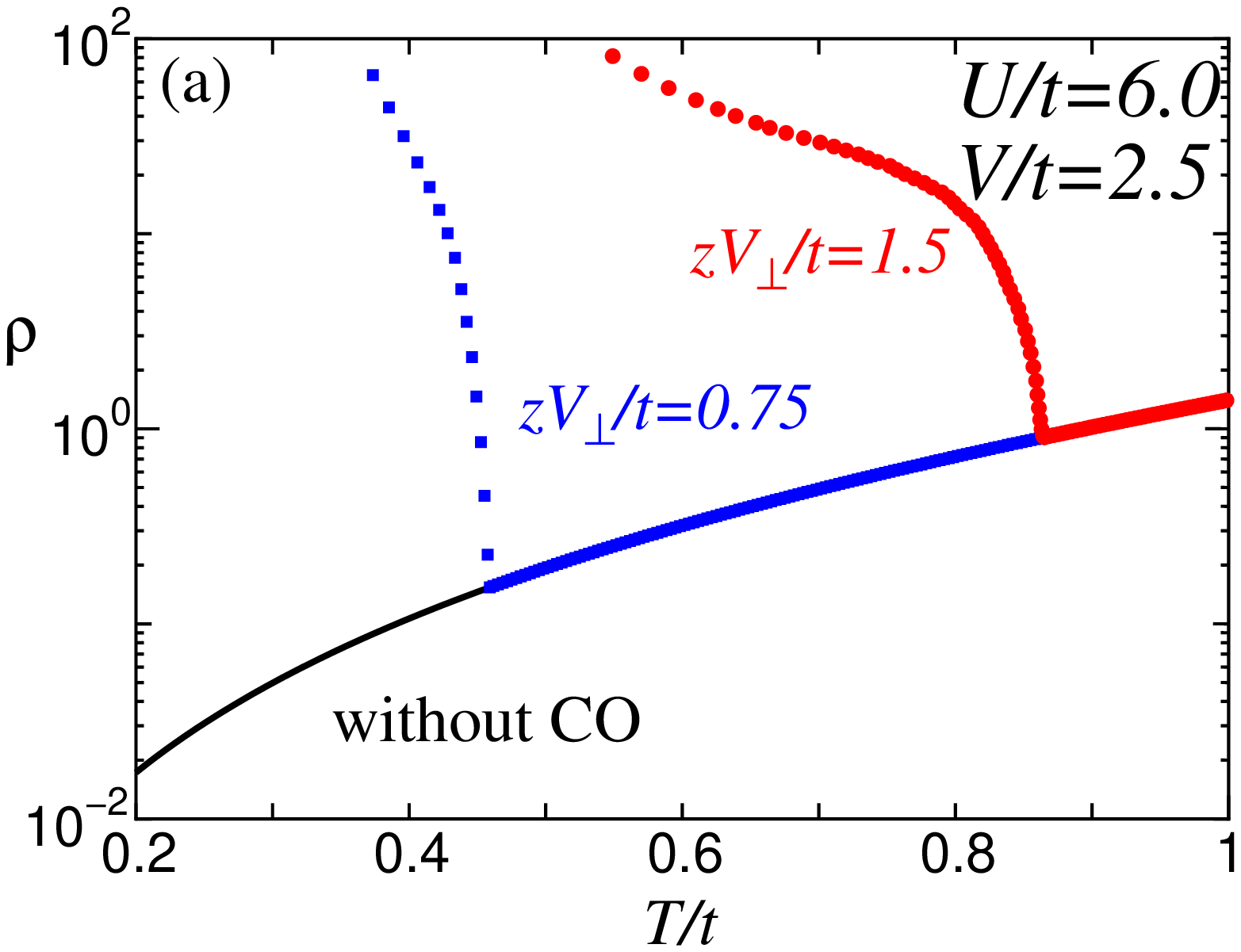} \\
\includegraphics[width=6cm]{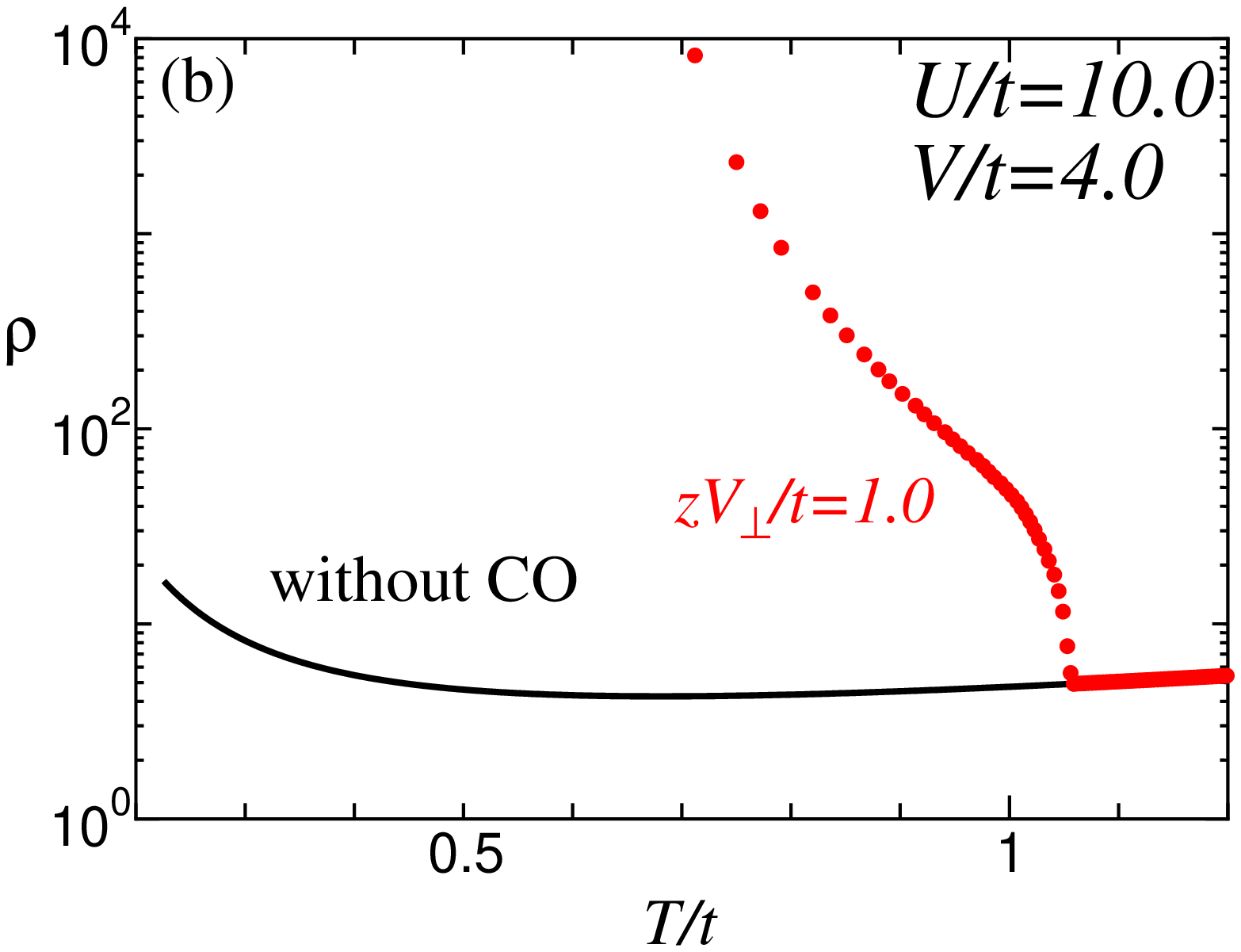}
\end{center}
\caption{
(Color online)
$T$ dependence of 
the resistivity $\rho(T)$ in arbitrary unit, 
(a) for the region (ii) and 
(b) for the region (iii) [same parameters as in Fig.~\ref{f3}].
The solid curve shows the data for $V_{\perp}=0$, 
 without charge order at finite $T$.
}
\label{f4}
\end{figure}

Next we discuss the $T$ dependence of the electrical resistivity $\rho(T)$.
Based on the memory function approach, \cite{Giamarchi1991,Giamarchi1992} 
 we perform the perturbation expansion 
 with respect to $g_{3 \perp}$ and $g_{1/4}$
in eq. (\ref{eqn:Hrho}) and then 
 the formula for the resistivity 
 is given by \cite{Tsuchiizu2001} 
\begin{align}
\rho(T) =& \frac{1}{\pi v_{\rho0}^2 \alpha} 
\Bigg[
g_{3\bot}^2
 \left( \frac{2\alpha}{\xi_T}\right)^{4 K_{\rho0} -3} 
B^2(K_{\rho0},K_{\rho0}) \nonumber \\
&
 + g_{1/4}^2 \left( \frac{2\alpha}{\xi_T}\right)^{16 K_{\rho0} -3}
B^2(4K_{\rho0},4K_{\rho0}) 
\Bigg],
\label{eqn:RT}
\end{align}
where $B(x,y) = \Gamma(x) \Gamma(y)/\Gamma(x+y)$ 
is the beta function and
$\xi_T=v_{\rho0}/(\pi T)$ is the thermal coherence length.
Here we note that the anomalous power-law behavior 
 $G_{3 \bot}^2 T^{4K_\rho -4}$ and 
 $G_{1/4}^2 T^{16K_\rho -4}$ can already be seen from 
 the perturbationaly-obtained form, eq. (\ref{eqn:RT}). 
However this expression is valid only in the high-$T$ region.
In order to examine qualitative behaviors in the wide $T$ range,
 we reinforce these expressions within the RG framework.
 \cite{Giamarchi1991,Giamarchi1992}   
By noting that the scaling dimension of the $4k_F$ and $8k_F$ umklapp
scatterings are $2K_{\rho0}$ and $8K_{\rho0}$, 
  the couplings are
  scaled (at tree level) as
  $G_{3 \bot} (l) = G_{3 \bot}(0) \exp \left[(2 - 2 K_{\rho0}) l\right]$
 and 
  $G_{1/4} (l) = G_{1/4}(0) \exp \left[(2 - 8 K_{\rho0}) l\right]$.
By inserting these relations into eq. (\ref{eqn:RT}), 
 and replacing the bare parameter $K_{\rho0}$ by the 
 renormalized parameter $K_{\rho}(l)$, 
 we obtain the formula improved by the RG method. 
Then the $T$ dependence can be 
 computed as 
\begin{align}
\rho(T) =&  
 \frac{2 \pi^2  T}{v_{\rho 0}} 
\Bigg[
G_{3 \bot}^2(T) 
B^2(K_\rho(T),K_\rho(T)) \nonumber \\
&\qquad
 + 4 G_{1/4}^2(T) 
B^2(4K_\rho(T),4K_\rho(T)) 
\Bigg],
\label{eqn:RT_RG}
\end{align}
where the $T$-dependent coupling constants
 are obtained by the procedure in \S~\ref{sec.Formulation}. 

In Fig. \ref{f4}, 
we show the results of $\rho(T)$ thus calculated, 
for the regions (ii) and (iii).
First, 
in the case of $V_\perp=0$ without CO at finite-$T$,  
 the system shows a metallic behavior for the whole $T$ range 
 for region (ii), i.e., $\rho(T)$ decreases with decreasing $T$, 
 whereas for region (iii), 
 it is insulating below the $T$ scale of the charge gap,  
 i.e., $\rho(T)$ increases with decreasing $T$. 
This behavior reflects the ground-state properties 
 in the absence of the interchain coupling, i.e.,
 the TL liquid in the region (ii), while 
 the CO insulating state in the region (iii). 
A noticeable point is that 
 $\rho(T)$ shows insulating behavior 
 even without long range order of CO~\cite{Tsuchiizu2001} 
 due to the low-dimensional fluctuation effect. 

When $V_\perp \neq 0$, $T_{\rm CO}$ becomes finite, 
 then one can see a clear cusp in $\rho(T)$ at $T=T_{\rm CO}$. 
Just below $T_{\rm CO}$, 
 $\rho(T)$ shows a curve which is convex upward in the semi-log plot, 
 reflecting the gap opening with the rapid growth of the CO order parameter $n$.
At lower $T$, 
 the curve turns convex downward, since at sufficiently low $T$, 
 $n$ becomes almost $T$ independent and therefore the gap can be considered 
 as a constant value, $\Delta$; then 
 an activation-type behavior $\rho(T) \propto \exp{(\Delta/T)}$ is expected.
Such a behavior is common for all the parameters we have considered, 
 as shown in Fig.~3. 
The abrupt change at $T=T_{\rm CO}$ 
 is indeed due to the emergence of 4$\kf$-umklapp
 scattering which originates from the gap in the energy dispersion at $\pm
 2 \kf$ owing to CO. 
This is in clear contrast with the behavior of the spin susceptibility $\chi_\sigma (T)$ 
 shown in Fig. 2 with only a tiny singularity at the transition.

\section{Summary and Discussions}
\label{sec.Discussions}

In the present paper, we have formulated a new theoretical 
framework to investigate 
 the finite-$T$ properties of Q1D electron systems. 
The method has been applied to the CO phase transition 
 in the Q1D EHM at quarter filling, 
 where extended Hubbard chains are coupled via 
 interchain Coulomb repulsion treated
 within the interchain mean-field approximation.
 
In our scheme, 
 we derive the bosonized Hamiltonian for the effective 1D model 
 and treat the RG equations, and by stopping the scaling procedure 
 at the corresponding scale 
 we obtain finite-$T$ properties of the system. 
As for the initial values of the RG equations 
 we use the numerical results for the small systems obtained by the Lanczos
 ED method; 
 these provide quantitatively good estimates 
 even for strong coupling regime, 
 in contrast with the conventional treatment where 
 only the interaction processes between electrons near the Fermi energy 
 are taken into account. 
In addition, 
QMC method is employed to calculate $T$ dependence of the CO order
parameter, which is necessary for the quantitative calculations below
$T_{\rm CO}$ but difficult to determine in the bosonization + RG scheme.   
This framework enables us to calculate physical quantities across
$T_{\rm CO}$
such as the spin susceptibility $\chi_\sigma (T)$ as well as the
electrical resistivity $\rho(T)$
which is hard to calculate by the  QMC simulation. 
 
The results show that, CO leads to an enhancement of $\chi_\sigma (T)$,
 mainly due to the reduction of the spin velocity, but without any steep
 singularity at $T=T_{\rm CO}$.  Such features in the $\chi_\sigma
 (T)$ curves shown in Fig. 2 are consistent with those calculated by purely
 numerical methods in refs.~\citen{Seo2007} and \citen{Otsuka2008}, which also
 showed a slight enhancement below $T_{\rm CO}$.  
In addition, our results
 share similarities with the $T=0$ properties of the 1D EHM, where
 $\chi_\sigma(T=0)$ shows no singular behavior at the critical value of
 $V$ for the emergence of CO, and continuously enhances when entering
 the CO phase~\cite{Tanaka2005}.  On the other hand, as for $\rho(T)$,
 the CO phase transition results in a sudden increase with a change of
 the slope with decreasing $T$, due to the generation of 4$\kf$-umklapp
 scattering which originates from the gap formation in the energy
 dispersion at $\pm 2 \kf$.  This is the first theoretical work, to the
 authors' knowledge, calculating $\rho(T)$ across the CO transition
 temperature starting from a microscopic correlated electronic model and
 taking full account of the quantum and thermal fluctuation effects.

Such results can be compared with the experiments, where $\chi_\sigma
(T)$ and $\rho(T)$ are the most essential information for the bulk
magnetic and electric properties of the system, measured most commonly.
In fact, the sudden increase and the change of slope in $\rho(T)$ are
observed in a wide classes of compounds showing CO, even in materials
which are apparently not applicable to our Q1D model.  On the other
hand, there are differences in the behavior of $\chi_\sigma (T)$ at
$T_{\rm CO}$ among the compounds, as discussed below.  Nevertheless, in
many materials no noticeable change is seen in $\chi_\sigma (T)$,
indicating a transition to the paramagnetic insulating CO phase, as in
our calculations.

For example, in quarter-filled molecular conductors where many compounds
show CO, Q1D materials such as (TMTTF)$_2X$ and (DCNQI)$_2X$ (except the
$\pi$-d mixed compound $X=$Cu) are candidates to be directly compared to
our results, since ours can be considered as a microscopic model for
their electronic properties.  Several members of the (TMTTF)$_2X$ family
showing CO, such as $X$=SbF$_6$, AsF$_6$, and ReO$_4$, indeed show kinks
at $T_{\rm CO}$ in their transport properties~\cite{Coulon1985,Nad2006}.
A difference between our model and the situation in the actual TMTTF
compounds is that, the present calculation does not include the
intrinsic lattice dimerization along the chains while it exists in the
materials, which leads to another non-linear term in the bosonized
Hamiltonian~\cite{Tsuchiizu2001}.  In (DI-DCNQI)$_2$Ag, even though
recent studies~\cite{Kakiuchi2007,Seo2009} revealed that the system
undergoes a more complex charge-lattice ordering than the simple CO we
investigate in this paper, the $T$ dependence in $\rho(T)$ across
$T=T_{\rm CO}$ observed at high pressure resembles our calculated data,
whereas the kink at $T=T_{\rm CO}$ is smeared out at ambient
pressure~\cite{Hiraki1996,Itou2004}.  All these Q1D molecular conductors
show no anomaly in the bulk $\chi_\sigma (T)$ at $T=T_{\rm CO}$, while NMR
measurements show the appearance of atomic sites showing different
Knight shifts and relaxation rates~\cite{Hiraki1998,Chow2000}: these are
also consistent with our analysis.  Recently, several Q1D compounds
without dimerization have been synthesized where a CO transition is
suggested, such as ({\it o}-DMTTF)$_2$Br~\cite{Fourmigue2008} and
(EDT-TTF-CONMe)$_2X$ [$X$= AsF$_6$ and
Br]~\cite{Heuze2003,Auban-Senzier2009}.  There, as in (DI-DCNQI)$_2$Ag
at ambient pressure, the anomaly at $T=T_{\rm CO}$ in $\rho(T)$ is
not clearly seen possibly due to the strong fluctuation which may be
underestimated in our calculation due to the interchain mean-field
treatment.  A noticeable point is that in ({\it
o}-DMTTF)$_2$Br~\cite{Fourmigue2008}, $\chi_\sigma (T)$ shows a steep
decrease at around $T=T_{\rm CO}$, distinct from the other Q1D materials
above with a smooth variation there, whose origin remains unclear.

Now there are many quasi-two-dimensional (Q2D) molecular conductors
found to show CO.  In such cases, the anomaly in $\rho(T)$ and the
characteristic curvature below $T=T_{\rm CO}$ are again ubiquitously observed,
while $\chi_\sigma (T)$ show different behaviors from material to
material.  
The latter diversity is due to the fact that, in the Q2D
compounds, there exists a variety in the anisotropy of transfer
integrals originated from different molecular packings, which results in
different anisotropic exchange couplings connecting the localized spins
when CO is formed: For example, a spin gapped behavior is observed in
$\alpha$-(BEDT-TTF)$_2$I$_3$~\cite{Rothaemel1986} and
$\beta''$-(DODHT)$_2$PF$_6$~\cite{Nishikawa2005} due to the alternation
in the exchange couplings along the charge rich sites.  On the other
hand, behavior analogous to our calculation is seen in
$\theta$-(BDT-TTP)$_2$Cu(NCS)$_2$~\cite{Ouyang2001,Yakushi2002}, where
the $\rho(T)$ curve shows a rapid increase at around $T_{\rm CO}=250$ K,
where $\chi_\sigma (T)$ shows no anomaly and below which the system is
paramagnetic.  
In $\theta$-(BEDT-TTF)$_2$RbZn(SCN)$_4$, where the
molecular packing in the 2D plane is the same, $\chi_\sigma (T)$ across
$T_{\rm CO}=190$ K is also paramagnetic; however, in $\rho(T)$ a large
jump is observed at $T=T_{\rm CO}$~\cite{Mori1998}.  The difference
between the two $\theta$-type compounds is in their CO pattern: the
`vertical stripe' in the former and the `horizontal stripe' in the
latter.  This results in different manners in coupling to the lattice
degree of freedom, leading to the second-order vs strong first-order
nature of the CO phase transition.  We note that most of the Q2D
compounds show a second-order or a weak first-order phase transition,
and $\theta$-(BEDT-TTF)$_2$RbZn(SCN)$_4$ is rather exceptional.  Another
point we note here is that under pressure, in several Q2D compounds such
as $\alpha$-(BEDT-TTF)$_2$I$_3$~\cite{Tajima2000} and $\beta$-({\it
meso}-DMBEDT-TTF)$_2$PF$_6$~\cite{Kimura2004,Morinaka2009}, $\rho(T)$
curve at low $T$ points toward a finite value extrapolated to $T=0$; it
does not diverge as in our calculation.  Such a behavior is observed
near the border between CO and uniform metallic phase without CO, and
suggests the existence of a CO metallic phase.  This is not realized in
our calculation, where the CO phase is always insulating, therefore can
be interpreted as the dimensionality effect in the transfer integrals.
In fact, the CO metallic phase is suggested in calculations on the 2D
EHM~\cite{Seo2006}.

Many transition metal oxides, even with Q2D or three-dimensional
structures, show CO when the number of carriers become a fraction of the
lattice sites.  In such cases, again the sharp kink structure in
$\rho(T)$ is widely observed, e.g., in Nickelates, Manganites,
Vanadates, and Iron based compounds~\cite{Imada1998}.  However, the
measurements for $\chi_\sigma (T)$ shows a variety in their behavior,
which is due to the same origin as in the Q2D molecular materials
mentioned above: the variety in the anisotropy of the exchange
couplings.  One recent example where $\chi_\sigma (T)$ is continuous at
the CO transition with a slight increase below $T_{\rm CO}=130$~K is
$\beta$-Na$_{1/3}$V$_{2}$O$_5$~\cite{Yamada1999}.  This compound is a
Q1D compound but has a complicated crystal structure with a filling
factor of 1/12; nevertheless the behavior, in the $\rho(T)$ measurement
as well for different pressures~\cite{Yamauchi2008}, resembles our
calculations.  The results of calculation that we obtain can be applied
to many material systems.

\section*{Acknowledgment}

We thank T. Hikihara for valuable discussions on 
  the Lanczos numerical technique.
This work was supported by MEXT 
Grant-in-Aid for Scientific Research (20110002, 20110004, 
and 21740270) 
and by Nara Women's University Intramural Grant for Project Research.

\appendix

\section{%
Combined Approach of Numerical Analysis, Bosonization, and
 Renormalization Group Method}
\label{ap.combined_approach}

In this appendix we explain how to determine 
the $T$-dependent parameters 
 needed to evaluate  the formulae of eqs. (\ref{eqn:chis}) and (\ref{eqn:RT_RG}), 
 by combining the numerical results for the finite size system 
 and the RG equations.  

The TL parameters, $K_\rho^L, K_\sigma^L$, and 
 the velocities, $v_\rho^L, v_\sigma^L$,  
 for the charge and spin degrees of freedom, respectively, 
 in a finite $L$ sites chain 
 can be calculated exactly by the Lanczos ED technique 
 using several standard relations~\cite{Schulz1994,Giamarchi}.  
The quantities above (we omit the superscript $L$) can be expressed as 
\begin{subequations}
\begin{align}
K_\rho  =
\frac{1}{2} \left(\pi \kappa D_\rho \right)^{1/2},
&\quad
v_\rho 
=
\left(\frac{D_\rho}{\pi \kappa} \right)^{1/2},
\\
K_\sigma  =
\frac{1}{2} \left(\pi \chi_s D_\sigma \right)^{1/2},
&\quad
v_\sigma 
=
\left(\frac{D_\sigma}{\pi \chi_s} \right)^{1/2},
\end{align}%
\label{eq:app-formula}%
\end{subequations}
where $\kappa$ and $\chi_s$ are 
the compressibility and the spin susceptibility, respectively,  
whereas $D_\rho$ and $D_\sigma$ are 
the Drude weights for the charge and spin currents.  
For the finite size systems, $\kappa$ and $\chi_s$ are given by 
\begin{subequations}
\begin{align}
\kappa^{-1}  
=&
\frac{L}{4} 
\left[
E_0(N_\uparrow+1,N_\downarrow+1)+E_0(N_\uparrow+1,N_\downarrow-1)
\right. \nonumber\\ & {} \left.
  -2E_0(N_\uparrow,N_\downarrow)
\right] ,
\\
\chi_s^{-1}  
=&
2L 
\left[
E_0(N_\uparrow+1,N_\downarrow-1)
-E_0(N_\uparrow,N_\downarrow)
\right] ,
\end{align}%
\label{eq:app-kappachis}%
\end{subequations}
where 
 $E_0(N_\uparrow,N_\downarrow)$ is the ground-state energy of
 the system with 
 $N_\uparrow$ spin-up and $N_\downarrow$ spin-down electrons.
For the estimation of the Drude weights, we use the relations
\begin{equation}
D_\rho =
 \frac{\pi}{L} \left. 
 \frac{\partial^2 E_0(\phi)}{\partial \phi^2}
\right|_{\phi=0}
, \quad
D_\sigma =
 \frac{\pi}{L} \left. 
\frac{\partial^2 E_0(\phi')}{\partial \phi'^2}
\right|_{\phi'=0}, 
\label{eq:app-Drude}
\end{equation}
where $E_0(\phi)$ ($E_0(\phi')$) is the ground-state energy in the
presence of the charge (spin) flux $\Phi=L\phi$ ($\Phi'=L\phi'$) through
the 1D ring. 
Then 
  the TL liquid parameters and velocities 
  are calculated exactly 
  for finite size systems (up to 16 sites) 
 using the Lanczos algorithm.

To estimate 
the TL-liquid parameters at finite-$T$, 
 we extend a recently-developed method
 combining the RG method with numerical results
 on finite size systems 
 to study the charge degree of freedom in the 1D EHM at quarter-filling~\cite{Sano2004}. 
The essential effects of the nonlinear terms are to renormalize 
 the parameters, namely, the TL-liquid parameters have 
 explicit size dependences. 
In ref. \citen{Sano2004}, it has been assumed that 
 the $L$ dependence of $K_\rho$ 
 is governed by the RG scaling equations for the nonlinear term $G_{1/4}$, 
 and $K_\rho^L$ for several-system sizes $L$ 
 are used as the initial condition 
 where the data is fitted to the scaling equations using the relation $l=\ln L$.

By following this idea, we determine
    the initial conditions  of the RG equations 
 $\left\{ K_\rho(l_1),G_{3\perp}(l_1),G_{1/4}(l_1) \right\}$
     for the charge degree of freedom, 
     eqs.\ (\ref{eqn:RGC1})-(\ref{eqn:RGC3}), 
as follows.  
(i) We calculate $\kappa$ and $D_\rho$ 
 using eqs. (\ref{eq:app-kappachis}) and (\ref{eq:app-Drude}) 
 which give the TL-liquid parameter by eq. (\ref{eq:app-formula}), 
 for three kinds of system size,  
 $K_\rho^{L=8}$, $K_\rho^{L=12}$, and $K_\rho^{L=16}$.
(ii) We set  $K_\rho(l_1)=K_\rho^{L=8}$ and we tentatively substitute
some values to $G_{3\perp}(l_1)$ and $G_{1/4}(l_1)$.
  By solving eqs.\ (\ref{eqn:RGC1})-(\ref{eqn:RGC3}) with these
  tentative initial values, we 
  calculate $K_\rho(l_2)$ and $K_\rho(l_3)$ where 
  $l_2=\ln 12$, and $l_3=\ln 16$.
(iii) We estimate   $(\delta K_\rho)^2 = (K_\rho(l_2)-K_\rho^{L=12})^2 
  + (K_\rho(l_3)-K_\rho^{L=16})^2$. 
Through these steps, 
 we optimize the initial values of 
 $G_{3\perp}(l_1)$ and $G_{1/4}(l_1)$ 
 so as to minimize the quantity $\delta K_\rho$.
Here we have assumed that $G_{3\perp}(l_1)$ ($G_{1/4}(l_1)$) is 
  odd (even) as a function of $zV_\perp n$, by which 
the trivial limit that  $G_{3\perp}(l_1)$ should reduce to zero 
  for $zV_\perp n=0$ is satisfied automatically.
This assumption can be justified by considering the perturbative forms.

\begin{figure}[t]
\begin{center}
\includegraphics[width=6cm]{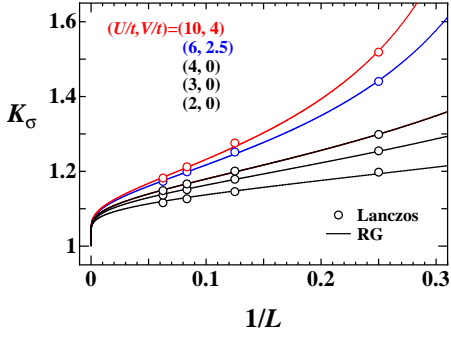}
\end{center}
\caption{
(Color online)
TL liquid parameter $K_\sigma$ 
 of finite size systems $L=4,8,12,16$ (open circles) 
 for several choices of $(U/t,V/t)$.
The solid lines denote  
 the scaling trajectories of $K_\sigma(l)$  with $l=\ln L$.
}
\label{fig:ksigma-traj}
\end{figure}

For the spin channel, 
  the initial value of $G_{1\bot}$ in eq. (\ref{eqn:RGS}) is determined from 
  the values of $K_\sigma$ of finite size systems.
Owing to the spin-rotational SU(2) symmetry, 
  the relation $K_\sigma=[(1+G_{1\bot})/(1-G_{1\bot})]^{1/2}$
  (eq.(\ref{eqn:SU22})) still
   holds even for the finite size systems.
By using this relation, 
  $G_{1\bot}$ for a finite size system is obtained from $K_\sigma$,
 and can be used as the initial condition.
This means that the initial condition of the nonlinear term  
 for the spin channel can be determined without the fitting procedure
  and we can directly check the assumption in ref. \citen{Sano2004}.
For the 1D EHM,
the TL liquid parameter $K_\sigma^L$ in the finite size system
  ($L=4,8,12,16$) and the scaling trajectory given by eq. (\ref{eqn:RGS})
 with $l=\ln L$
  are shown in Fig.\ \ref{fig:ksigma-traj}.
The initial value of $G_{1\bot}$
  is estimated with the least-square fit, i.e.,  
  in order to minimize 
 $(K_\sigma^{L=4}-K_\sigma(l_0))^2 
  +(K_\sigma^{L=8}-K_\sigma(l_1))^2
 + (K_\sigma^{L=12}-K_\sigma(l_2))^2
 +(K_\sigma^{L=16}-K_\sigma(l_3))^2$.
We can see that the scaling trajectory reproduces 
  the numerical ED data very well, even for small system sizes.
We note that the downward-convex dependence at large $1/L$ cannot be 
   obtained by the one-loop RG, while
the upward-convex $1/L$ dependence
  at small $1/L$ is nothing but the logarithmic singularity 
\cite{Cardy}
and
can be reproduced even in the one-loop level.

The $T$ dependence of the $G_{4\sigma}(T)$ coupling
  can be obtained by using  eq. (\ref{eqn:g4s}), 
where  we need the $T$ dependence 
  of the spin velocity.
The RG equation for the spin velocity has not been obtained without
ambiguity, since 
   the correction is not  logarithmic
  and the scaling invariance is not retained.
Thus we utilize the simple size dependence of the velocity
  to examine its $T$ dependence.  
The finite size scaling relation  is expressed as 
\begin{align}
 v_\sigma (L) = v_\sigma (\infty) + c_1/L + c_2/L^2,   
\label{eqn:vsFSS}
\end{align} 
where $c_1$ and $c_2$ are numerical constants. 
In order to obtain the $T$ dependence of the spin velocity, 
we use eq. (\ref{eqn:vsFSS}) by substituting 
$ L =\mathrm{e}^l \simeq Ct/T$ with $C$
 being an $O(1)$ numerical constant.

\section{Spin Susceptibility in One-Dimensional Hubbard Model}
\label{ap.susceptibility}

\begin{figure}[t]
\begin{center}
\includegraphics[width=6.5cm]{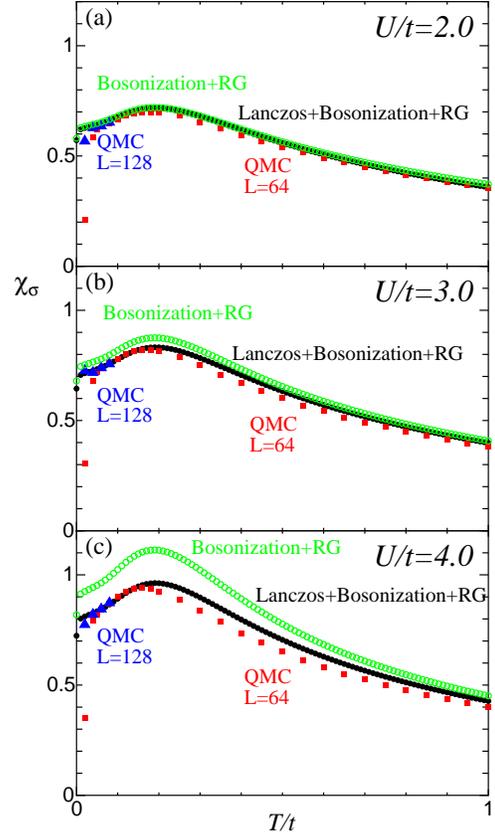}
\end{center}
\caption{
(Color online)
$T$ dependence of 
 the uniform spin susceptibility $\chi_\sigma$ in unit of $1/(ta)$ 
 for the 1D Hubbard model,  
 for several choices of $U/t$. 
The filled circle and the open circle represent the results 
 by ``Lanczos+Bosonization+RG'' 
 and 
 those by ``Bosonization+RG'',
 respectively (see text). 
These results are compared with QMC results (filled square
 for $L=64$ and filled triangle for $L=128$).   
}
\label{fA1}
\end{figure}

\begin{figure}[t]
\begin{center}
\includegraphics[width=6cm]{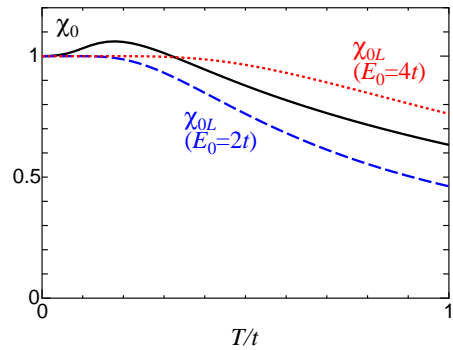}
\end{center}
\caption{
(Color online)
$T$ dependence of 
 the normalized uniform susceptibility $\chi_0(T)$ 
 in the case of $\epsilon(K) = -2t \cos Ka$ (solid curve) 
 and 
 the susceptibility $\chi_{0L}(T)$ for $\epsilon(K) = v_F (pK - k_F)$ 
 with  $|\epsilon(K)| < E_0/2$ 
 (dashed curve for $E_0 = 2t$ and dotted one for $E_0 = 4t$). 
}
\label{fA2}
\end{figure}

In this section, 
we demonstrate the validity and the advantage 
of our combined analytical and numerical method 
 by evaluating the uniform spin susceptibility 
of the 1D Hubbard model as a function of $T$. 
The present method 
(Lanczos+Bosonization+RG) is compared with 
the QMC method,  
as well as the conventional method
   based on the RG theory (Bosonization+RG). 
We summarize the results in Fig. \ref{fA1}, where 
 the case for $U/t=2.0$ (a), 3.0 (b), and 4.0 (c) are shown.
Both the  ``Lanczos+Bosonization+RG'' method and the usual
``Bosonization+RG'' method can reproduce the QMC results
qualitatively. 
Especially, the present method gives more accurate results compared with
the usual method even in the strong interaction cases. 
We note that, the discrepancy between the present results
and the  QMC method is mainly due to the formula of the spin
susceptibility, eq. (\ref{eqn:chis}). 
The formula (\ref{eqn:chis}) 
 is based on the RPA formalism with renormalized parameters, 
which overestimates the absolute value.\cite{Fuseya2005} 
It should be noted that 
the noninteracting susceptibility $\chi_0(T)$ 
  is evaluated by using the cosine-band dispersion  even in the
``Bosonization+RG'' method instead of the linearized
dispersion, in order to see the importance of choice of the initial
values of the RG equations. 
%
%
In the case of  the linearized dispersion, 
the noninteracting susceptibility is written as 
\begin{align}
 \chi_{0L}(T) = \tanh \frac{E_0}{4T},
\end{align}
where the energy dispersion is terminated as 
$-E_0/2 < v_F(pK-k_F) < E_0/2$ with $p=\pm$.
This shows monotonic decrease with increasing $T$ 
and the 
 comparison between the cosine-band dispersion and  the linearized dispersion
  is shown in Fig. \ref{fA2}.

\bibliographystyle{jpsj}
\bibliography{refs}

\begin{thebibliography}{10}
\bibitem{note}
In eq. (14) of ref. 12, $n$ is proproportional to the expectation value of
  $\cos{2\theta}$. The ambiguity enters in the coefficient $c_2$, of which its
  non-perturpative expression is unknown.
\end{thebibliography}
\end{document}